# Genetic Programming Framework for Fingerprint Matching

Ismail A. Ismail[*1], Nabawia A. ElRamly [#2], Mohammed A. Abd-ElWahid [#3], Passent M. ElKafrawy [#4] and Mohammed M. Nasef [#5]

[#] Department of Mathematics, Faculty of Science, Menoufia University, Egypt
[*] Dean, Faculty of Computers and informatics, Misr International University, Egypt
e.mail: mnasef81@yahoo.com

***Abstract*—** A fingerprint matching is a very difficult problem. Minutiae-based-matching is the most popular and widely used technique for fingerprint matching. The minutiae points considered in automatic identification systems are based normally on termination and bifurcation points. In this paper we propose a new technique for fingerprint matching using minutiae points and genetic programming. The goal of this paper is extracting the mathematical formula that defines the minutiae points.

***Index Terms***— Fingerprint matching, minutiae points, genetic programming

## I. INTRODUCTION

Fingerprint matching depends on the comparison of the characteristic of local ridges and their relationships. Widely used local ridge's characteristics, called minutiae in automatic fingerprint identification systems, are ridge termination and bifurcation [1], [2]. Most of the existing automatic fingerprint verification systems are based on minutiae features (ridge bifurcation and ending). Such systems first detect the minutiae in a fingerprint image and then match the input minutiae set with the stored template [3], [4], [5]. Extracting minutiae from fingerprint images is one of the most important steps in automatic fingerprint identification system. Because minutiae matching are certainly the most-well-known and widely used method for fingerprint matching [6], [7]. In this paper we use genetic programming (GP) to extract mathematical formulas for minutiae points (end points and bifurcation points). In section II we introduced genetic programming, and the effect of parameters on genetic programming, In section III we will explain the proposed methodology with experimental results. Finally, section IV provides a conclusion and some future work.

## II. GENETIC PROGRAMMING

Genetic programming (GP) is used for automated learning of computer programs. GP's learning algorithm is inspired by the theory of evolution and our contemporary understanding of biology and natural evolution. The most commonly used representation in genetic programming is the program tree. GP trees and their corresponding expressions can equivalently be represented in prefix notation (e.g., as Lisp expressions) [8], [9].

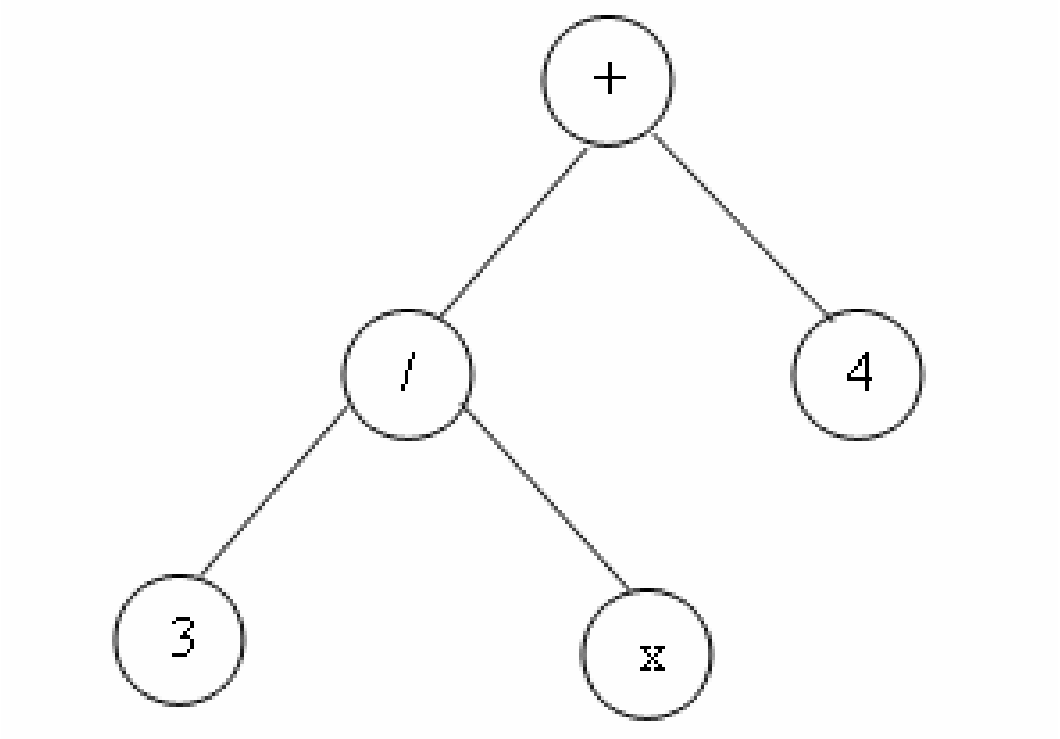

**Fig. 1**: Basic tree-like program representation used in GP

When applying genetic programming to a problem, there are five major preparatory steps involved [9], [10]. The following five parameters need to be determined before applying GP:
1) The set of terminals (e.g., the independent variables of the problem, zero-argument functions, and random constant) for each branch of the to-be-evolved program.
2) The set of primitive functions (e.g., Boolean functions, arithmetic functions, conditional functions) for each branch of the to-be-evolved program.

3) The fitness measure (for explicitly or implicitly measuring the fitness of individuals in the population).
4) Certain parameters for controlling the run, and
5) The termination criterion and method for designating the result of the run. Genetic operators.
The two main genetic operators are mutation and crossover [8], [10], [11]. Mutation works as follows: (i) randomly select a node within the parent tree as the mutation point; (ii) generate a new tree of maximum depth; and (iii) replace the subtree rooted at the selected node with the generated tree. For illustration refer to the mutation process in figure 2.
Crossover works as follows: (i) randomly select a node within each tree as crossover points, (ii) take the sub tree rooted at the selected node in the second parent and use it to replace the sub tree rooted at the selected node in the first parent to generate a child (and optionally do the reverse to obtain a second child). The crossover procedure is illustrated in figure 3.

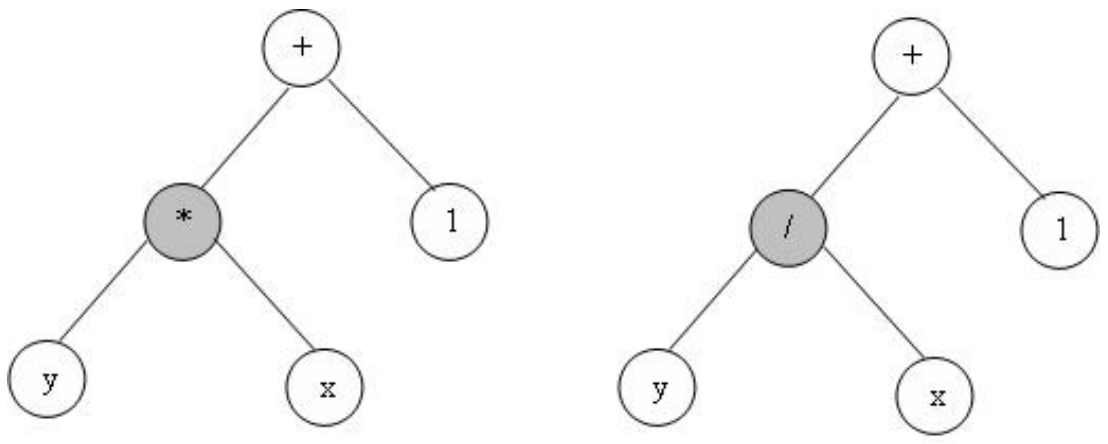

(a) Before Mutation (b) After Mutation

**Fig. 2**: Mutation in genetic programming

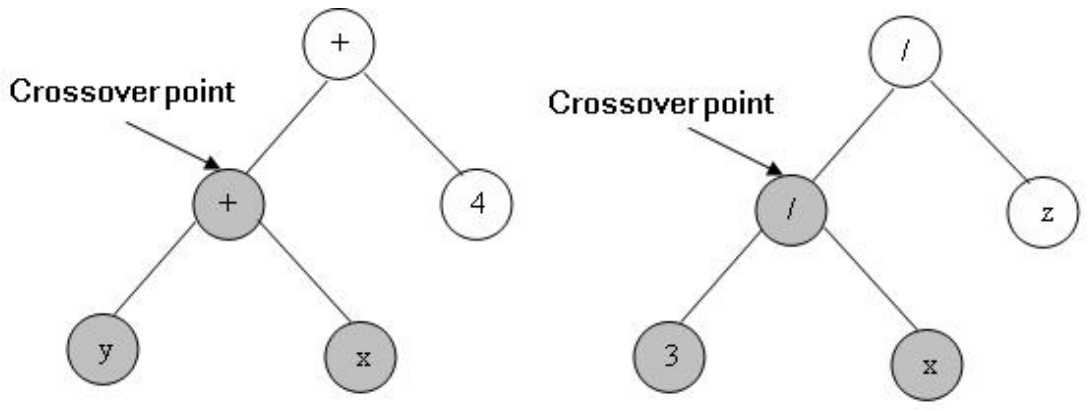

(a) Before Crossover

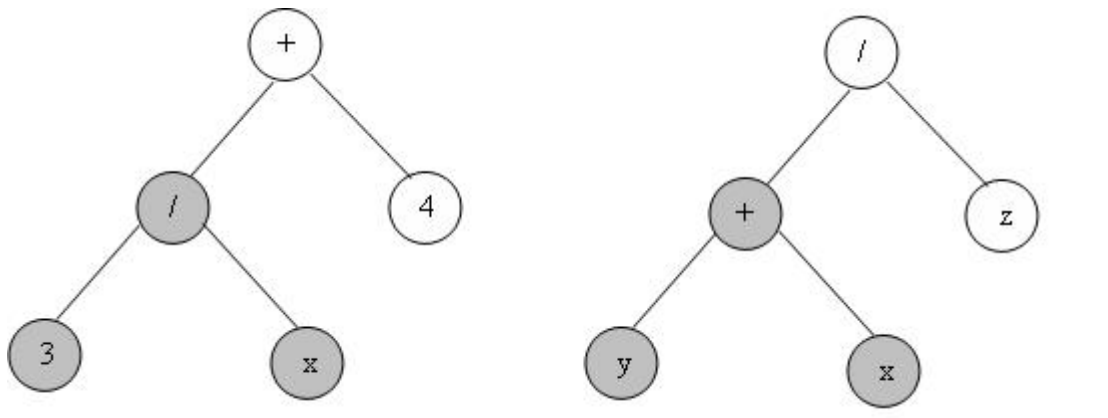

(b) After Crossover

**Fig. 3**: Crossover in genetic programming

In summary, genetic programming creates computer programs by executing the following steps, refer to [8], [10], [11].
Step 1 Assign the maximum number of generations to run and probabilities for cloning, crossover and mutation.
Step 2 Generate an initial population of computer programs of size N by combining randomly selected functions and terminals.
Step 3 Execute each computer program in the population and calculate its fitness with an appropriate fitness function. Designate the best-so-far individual as the result of the run.
Step 4 With the assigned probabilities, select a genetic operator to perform cloning, crossover or mutation.
Step 5 If the cloning operator is applied, then select one computer program from the current population of programs and copy it into a new population.

- If the crossover operator is applied, then select a pair of computer programs from the current population creates a pair of offspring programs and places them into the new population.
- If the mutation operator is applied, select one computer program from the current population, perform mutation and place the mutant into the new population.

Step 6 Repeat Step 4 until the size of the new population of computer programs becomes equal to the size of the initial population, N.
Step 7 Replace the current (parent) population with the new (offspring) population.
Step 8 Go to Step 3 and repeat the process until the termination criterion is satisfied.

## III. Proposed Methodology

To match a query fingerprint to another one, say matching fingerprint, we extract the minutiae points for the two fingerprints first. The next step is to obtain the mathematical formula for the query fingerprint using GP. Finally, we apply the minutiae points of the matching fingerprint into the mathematical formula. If the resulting y of these points is the same as of the query fingerprint then the two fingerprints are a match, otherwise, there is no match. In this section we explain our proposed methodology on a given fingerprint as an example. Our proposed methodology is divided into two parts. In the first part, we extract minutiae points. In the second part we use genetic programming to extract mathematical formulas describing the minutiae points (end points and bifurcation points) which is explained in full details in the next two subsections.

### *A. Extracting Minutiae Points*

In the first step the fingerprint image is to be enhanced and filtered using any enhancing and filtering technique. After that we extract the minutiae points from the enhanced image. Figure 4(a) shows the query fingerprint, figure 4(b) is the fingerprint after enhancement and filtering, figure 4(c) is the fingerprint image with the determined minutiae points shown in the figure. We used the crossing number (CN) method to perform the minutiae points extraction. This method extracts the ridge endings and bifurcation from the skeleton image by examining the local neighborhood of ridge pixel using a 3*3 window. After using CN method we can draw two graphs describing these points; see figure 5 where the two graphs presenting the minutiae points – end points and bifurcation points respectively. From the two graphs we can extract all data for minutiae points (end points and bifurcation points) as summarized in table I and II. Ending points were determined by three variables x, y, and angle, but bifurcation points were determined by five variables x, y, angle1, angle2, and angle3. To apply this notation to genetic programming, we will define the following for end points:

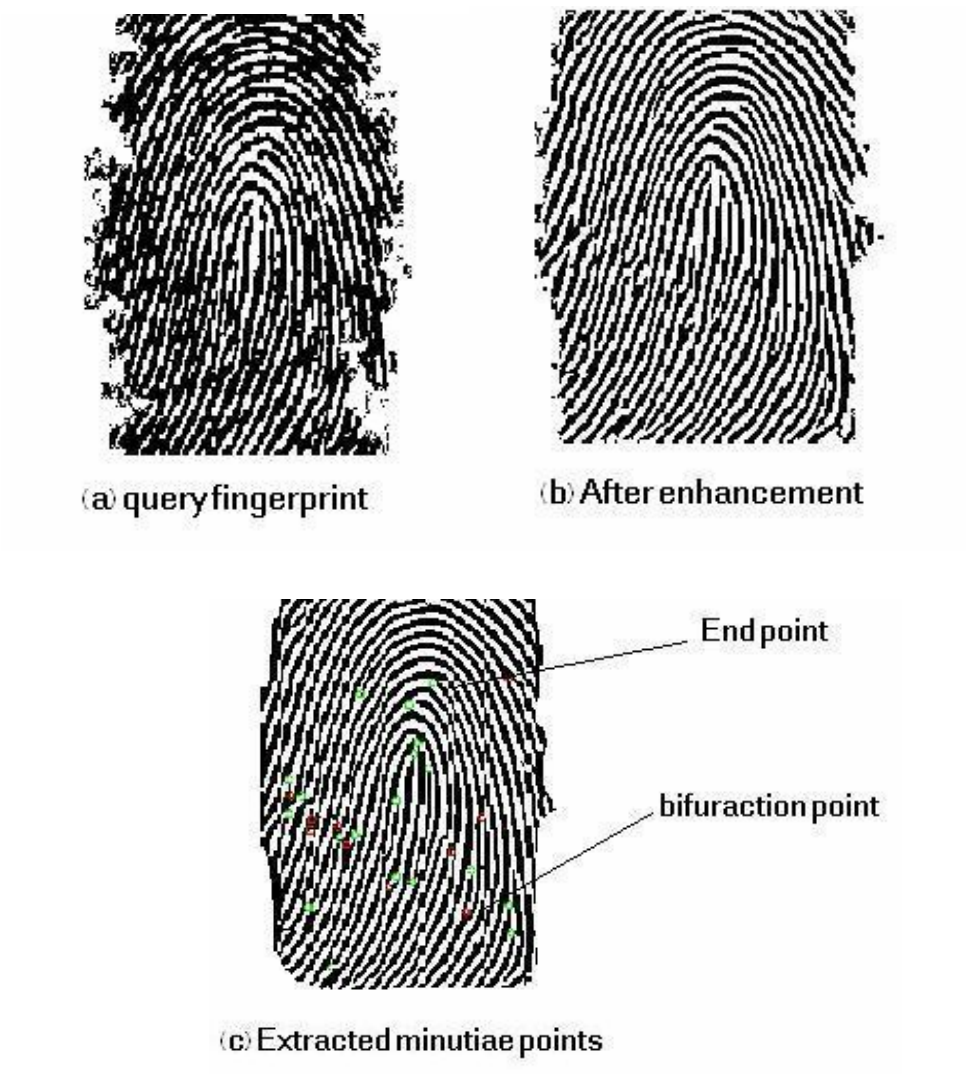

**Fig. 4**: Query fingerprint image, after enhancement and extracted minutiae points

- The terminal set for end points (x, y, angle, and random integer numbers).
- The function set for end points (+, -, *, and /).

On the other hand, we'll define the following for bifurcation points:

- The terminal set for bifurcation points (x, y, angle1, angle2, angle3, and random integer numbers).
- The function set for bifurcation points (+, -, *, and /).

Those points where defined in table I and II for the fingerprint in figure 4.

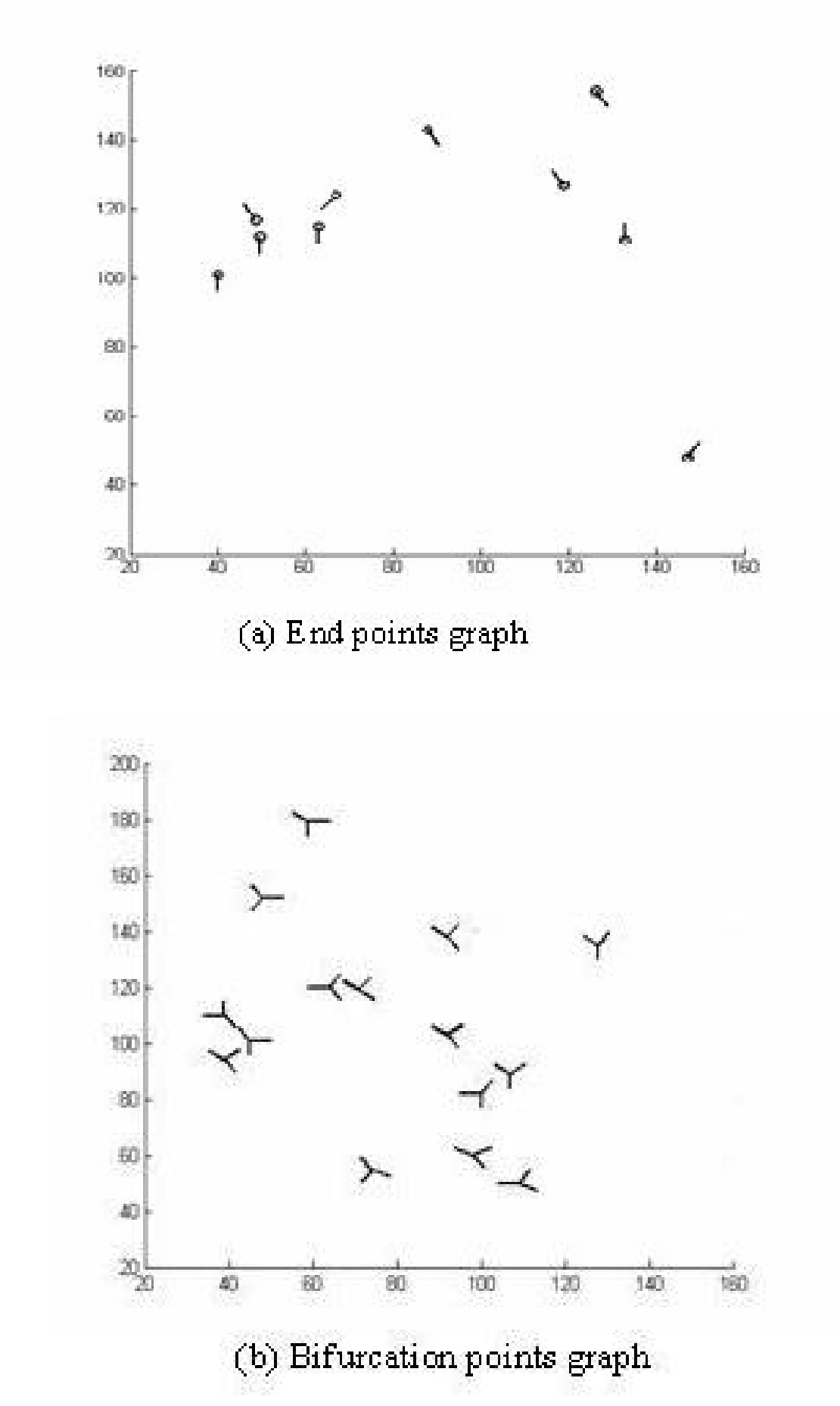

**Fig. 5**: Two graphs explain minutiae points

**TABLE I**: End Points Data

| x | angle | y |
|---|---|---|
| 147 | -1.05 | 48 |
| 40 | 1.57 | 101 |
| 133 | -1.57 | 111 |
| 50 | 1.57 | 112 |
| 63 | 1.57 | 115 |
| 49 | -2.09 | 117 |
| 67 | 2.36 | 124 |
| 119 | -2.09 | 127 |
| 88 | 1.05 | 143 |
| 126 | 1.05 | 154 |

**TABLE II**: Bifurcation points Data

| x | angle1 | angle2 | angle3 | y |
|---|---|---|---|---|
| 109 | 3.14 | -1.05 | .52 | 50 |
| 74 | 2.09 | -2.09 | .52 | 55 |
| 98 | -2.62 | 1.05 | -.52 | 60 |
| 100 | 3.14 | 1.57 | -1.05 | 82 |
| 107 | -2.36 | 1.57 | -.79 | 89 |
| 39 | -2.36 | 1.05 | -.79 | 94 |
| 45 | -2.09 | 1.57 | 0 | 101 |
| 92 | -2.36 | 1.05 | -.79 | 103 |
| 39 | 3.14 | -1.57 | 1.05 | 110 |
| 71 | -2.36 | -1.05 | .79 | 119 |
| 64 | 3.14 | 1.05 | -1.05 | 120 |
| 128 | -2.36 | 1.05 | -1.05 | 135 |
| 92 | -2.36 | 1.05 | -1.05 | 138 |
| 48 | 2.09 | -2.09 | 0 | 152 |
| 59 | -2.36 | 1.57 | 0 | 179 |

### *B. Using Genetic Programming (GP) With Minutiae Points*

In the previous section we extracted the minutiae points (end points and bifurcation points). In this section we will use this information and develop two mathematical formulas describing the relationship between the minutiae points using genetic programming. As defined in section II we defined the five main initiative parameters, see table III, to create the program that defines the formulas for the minutiae points of the fingerprints.

**TABLE III**: Genetic programming parameters

| | |
|---|---|
| Maximum number of Generations | 1700 |
| Size of Population | 2500 |
| Maximum depth of new individuals | 6 |
| Maximum depth of new sub trees for mutants | 4 |
| Maximum depth of individuals after crossover | 18 |
| Fitness-proportionate reproduction fraction | 0.1 |
| Crossover at any point fraction | 0.2 |
| Crossover at function points fraction | 0.2 |
| Number of fitness cases (end points) | 10 |
| Number of fitness cases (bifurcation points) | 15 |
| Selection method | fitness- proportionate with- over-selection |
| Generation method | Full |

In table I and II the variable y is a depended variable on x and the angles variables. In other words, y is the required output from the formula defined by x and the angles for each of the sets of points that define the end points and the bifurcation points. The GP program produces the formulas that is define y by knowing x and the angles. Figure 6a shows part of the mathematical formula for end points in S-expression, and figure 6b shows part of the mathematical formula for bifurcation points after running the GP algorithm.

```
(+ (- (% (+ (% (% -8 X) (*(% (+ (% ANGLE X) -8) -2) ANGLE))
      (% (% ANGLE -1) (% (+ -9 ANGLE) (- 3 -2))))
     (- (+ (% (+ (+ 7 ANGLE) -2) X) (- -5 2))
      (- (% -5 (* X -4)) (+ ANGLE 9))))
   (- (*(% (- X ANGLE) (+ X - 1))
      (- (% (+ (+ X (+ -2 (- (% X ANGLE) ANGLE))) X) -6)
       (* (% (- X ANGLE) (+ (* ANGLE X) -1))
        (- (% (% (+ ANGLE X)
              (* (% (* (*10 (+ ANGLE (+ 1 -3)))
                 (% (*ANGLE ANGLE) 2))
               (+ (+ 9 -3)
                 (% (- (+ (*(+ X X) X)))))
.
.
.
```

(a) End Points formula

```
(+ (% (- (+ (% (*10 (% (% (*-9 ANGLE2) ANGLE3) -2)) (% ANGLE3 -1))
       (*(*(- ANGLE3 3) (+ ANGLE2 ANGLE1))
        (+ (- -7 ANGLE3) (- ANGLE2 3))))
     (+ (- (- 10 -4) (*-7 ANGLE3))
      (- 9
       (% (*(*6 -8) ANGLE3)
        (% (- (+ (% (*10 (% (% (*-9 ANGLE2) ANGLE3) -2))
                 (% ANGLE3 -1))
              (*(*(- ANGLE3 3) (+ ANGLE2 ANGLE1))
               (+ (- ANGLE2 3) (- ANGLE2 3))))
            (+ (- (- 10 -4) (*-7 ANGLE3))
             (% (- (% (* ANGLE3 -3)
                   (- (+ (% ANGLE1
                        (+ (- (- (+ ANGLE3
                                (% (- (% (- (+ (*(+ (% (+ (% X
                                                         3)
                                                       (% (% (% (- -5
                                                                  X)
                                                             (% ANGLE2
                                                                6))
                                                          (- 2
                                                             (- ANGLE1
                                                                -5)))
                                                         3))
                                                  -3)
                                                 2)))))))
.
.
.
```

(b) Bifurcation points formula

**Fig. 6**: Part of mathematical formula for minutiae points (S Expression)

After we get these formulas, we want to compare the query fingerprint with all stored fingerprints to find a match to the query fingerprint. Assume that we have three stored fingerprints with the minutiae points (end points and bifurcation points) extracted previously. After that we'll apply each of these points in the mathematical formulas of the query fingerprint. The points that solve for y and is equal to the result, y, of the query fingerprint is a matching image. The results must be equal in end points formula and bifurcation points formula, otherwise there is no match.

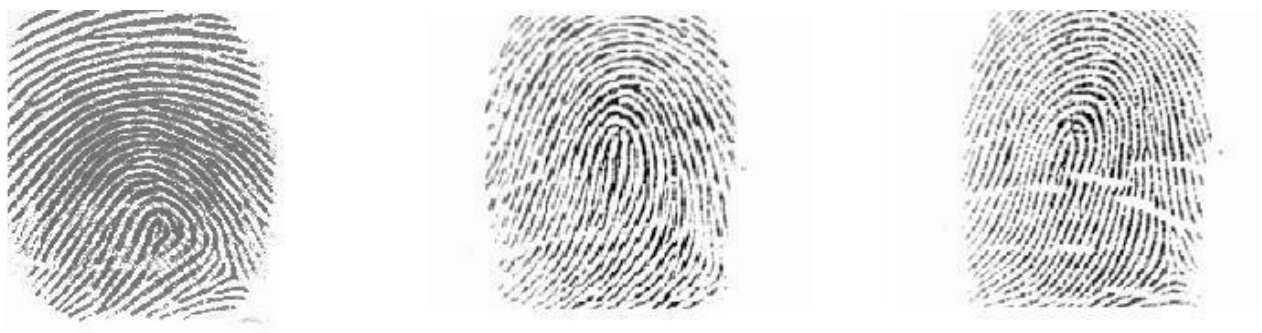

(a) image1 (b) image2 (c) image3

**Fig. 7**: Three different fingerprints

Table IV lists the end points of the fingerprints in figure 7. Actually, ten points are extracted for image 7a and image 7b while image 7c had fifteen end points extracted. Using mean square error method we found that image 7b matches the query image in end points but image 7a and image 7c are not a match, see table V. Tables VI, VII and VIII shows the bifurcation points of three fingerprints in figure 7. Table IX shows the results of evaluating these points in the mathematical formula of the query fingerprint (bifurcation mathematical formula). We can conclude that the number of bifurcation points is ten points in image 7a, fifteen bifurcation points in image 7b, and fourteen bifurcation points in image 7c. Using mean square error method we found that image 7b is a match to the query image in bifurcation points, whereas image 7a and image 7c are not a match. Finally, Fingerprint image 7b is matched with the query fingerprint image and fingerprints image 7a and image 7c are not.

**TABLE IV**: End points of three fingerprints.

| image1 | | image2 | | image3 | |
|---|---|---|---|---|---|
| x | angle | x | angle | x | angle |
| 86 | -2.62 | 147 | -1.05 | 104 | 3.14 |
| 158 | -.52 | 40 | 1.57 | 98 | 0 |
| 156 | -.52 | 133 | -1.57 | 121 | 2.09 |
| 93 | .52 | 50 | 1.57 | 65 | 2.36 |
| 111 | .79 | 63 | 1.57 | 130 | -2.09 |
| 24 | -2.36 | 49 | -2.09 | 133 | 1.57 |
| 112 | .52 | 67 | 2.36 | 88 | -2.09 |
| 161 | -2.09 | 119 | -2.09 | 107 | 1.05 |
| 103 | .52 | 88 | 1.05 | 128 | -1.57 |
| 151 | .52 | 126 | 1.05 | 144 | -1.57 |
| | | | | 69 | 1.57 |
| | | | | 99 | -2.62 |
| | | | | 73 | -2.09 |
| | | | | 62 | 0.79 |
| | | | | 120 | 2.62 |

**TABLE V**: The results of evaluating them in the mathematical formula of the query fingerprint (end points formula).

| image1 | image2 | image3 | query image |
|---|---|---|---|
| 111.62 | 48.00 | 150.47 | 48 |
| 96.73 | 101.00 | 224.74 | 101 |
| 96.62 | 111.00 | 149.19 | 111 |
| 160.05 | 112.00 | 122.94 | 112 |
| 153.90 | 115.00 | 129.03 | 115 |
| 98.99 | 117.00 | 150.43 | 117 |
| 172.66 | 124.00 | 121.71 | 124 |
| 134.89 | 127.00 | 152.10 | 127 |
| 166.66 | 143.00 | 109.83 | 143 |
| 199.05 | 154.00 | 112.52 | 154 |
| | | 117.86 | |
| | | 114.43 | |
| | | 119.19 | |
| | | 125.93 | |
| | | 129.56 | |

**TABLE VI**: Bifurcation points of image 1

| x | angel1 | angel2 | angel3 |
|---|---|---|---|
| 109 | 3.14 | .79 | -.52 |
| 96 | 2.62 | -2.09 | 0 |
| 149 | 2.62 | -1.57 | 0 |
| 110 | 2.62 | -2.09 | 0 |
| 122 | 2.62 | -1.57 | 0 |
| 80 | 3.14 | -1.57 | 1.05 |
| 116 | 2.36 | -2.62 | -.79 |
| 171 | 2.36 | -2.36 | -.79 |
| 154 | -2.62 | 1.57 | -1.05 |
| 167 | -2.62 | 1.05 | -.52 |

**TABLE VII**: Bifurcation points of image 2

| x | angel1 | angel2 | angel3 |
|---|---|---|---|
| 109 | 3.14 | -1.05 | .52 |
| 74 | 2.09 | -2.09 | .52 |
| 98 | -2.62 | 1.05 | -.52 |
| 100 | 3.14 | 1.57 | -1.05 |
| 107 | -2.36 | 1.57 | -.79 |
| 39 | -2.36 | 1.05 | -.79 |
| 45 | -2.09 | 1.57 | 0 |
| 92 | -2.36 | 1.05 | -.79 |
| 39 | 3.14 | -1.57 | 1.05 |
| 71 | -2.36 | -1.05 | .79 |
| 64 | 3.14 | 1.05 | -1.05 |
| 128 | -2.36 | 1.05 | -1.05 |
| 92 | -2.36 | 1.05 | -1.05 |
| 48 | 2.09 | -2.09 | 0 |
| 59 | -2.36 | 1.57 | 0 |

**TABLE VIII**: Bifurcation points of image 3
X Angle 1 Angle 2 Angle 3

| x | angel1 | angel2 | angel3 |
|---|---|---|---|
| 52 | 2.09 | -2.09 | .79 |
| 88 | -2.62 | 1.05 | 0 |
| 132 | -2.36 | 2.09 | -1.05 |
| 75 | -2.62 | 1.05 | -1.05 |
| 106 | -2.36 | 1.57 | -.79 |
| 123 | 2.09 | -1.57 | 1.05 |
| 115 | 3.14 | 1.05 | -.79 |
| 108 | -2.62 | -1.05 | .79 |
| 66 | -2.62 | 1.05 | -1.05 |
| 62 | -2.62 | -1.05 | .79 |
| 137 | 2.36 | .79 | -.79 |
| 77 | 2.09 | -2.09 | 0 |
| 75 | -2.09 | 1.57 | 0 |
| 78 | -2.62 | -1.05 | .79 |

**TABLE IX**: The results of evaluating the bifurcation formula of the query fingerprint on the 3 fingerprints.

| image 1 | image 2 | image 3 | query image |
|---|---|---|---|
| 97.06 | 50.00 | 70.79 | 50 |
| 157.71 | 55.00 | 178.81 | 55 |
| 153.76 | 60.00 | -341.05 | 60 |
| 155.82 | 82.00 | 182.66 | 82 |
| 156.58 | 89.00 | 89.06 | 89 |
| 134.07 | 94.00 | -81.82 | 94 |
| 325.16 | 101.00 | 98.94 | 101 |
| 314.5 | 103.00 | 115.07 | 103 |
| 227.62 | 110.00 | 177.93 | 110 |
| 65.89 | 119.00 | 115.95 | 119 |
| | 120.00 | 115.19 | 120 |
| | 135.00 | 149.14 | 135 |
| | 138.00 | 133.63 | 138 |
| | 152.00 | 115.70 | 152 |
| | 179.00 | | 179 |

## IV. CONCLUSION

In this paper we proposed a novel method for fingerprint matching. We used genetic programming with minutiae points to extract the mathematical formula that define fingerprint and can be used in matching between fingerprints. We can obtain theses mathematical formulas after enhancing and filtering the fingerprint after extracting the minutiae points in the fingerprint using crossing number (CN) method. Finally we can use all data about minutiae points (end points and bifurcation points) to extract the mathematical formula. In future work we try using genetic programming to classify fingerprints to decrease time search and match.